\documentclass[amsmath,amssymb,aps,prc,twocolumn,superscriptaddress,showpacs,preprintnumbers]{revtex4}

\usepackage{graphicx,color}
\usepackage{epstopdf}
\usepackage{multirow}
\usepackage{times}
\usepackage[normalem]{ulem}  

\renewcommand\sout{\bgroup \color{red} \ULdepth=-.5ex \ULset}
\newcommand{\NTP}{{N_\mathrm{TP}}}

\newcommand{\NMC}{{N_\mathrm{MC}}}
\newcommand{\ND}{{N_\mathrm{D}}}

\newcommand{\fW}{f_{\scriptscriptstyle{W}}}
\newcommand{\fH}{f_{\scriptscriptstyle{H}}}

\newcommand{\SHW}{{S_\mathrm{HW}}}
\newcommand{\SHWPA}{{S_\mathrm{HW}^\mathrm{(PA)}}}
\newcommand{\beq}{\begin{eqnarray}}
\newcommand{\eeq}{\end{eqnarray}}
\begin{document}
\preprint{KUNS-2616, YITP-16-40}

\title{
Entropy production from chaoticity in Yang-Mills field theory with use of 
the Husimi function
}

\author{Hidekazu Tsukiji}
\email[]{tsukiji@yukawa.kyoto-u.ac.jp}
\affiliation{Yukawa Institute for Theoretical Physics, Kyoto University,
Kyoto 606-8502, Japan}
\author{Hideaki Iida}
\thanks{The present address: Keio University}
\affiliation{Department of Physics, Faculty of Science, Kyoto University,
Kyoto 606-8502, Japan}
\author{Teiji Kunihiro}
\affiliation{Department of Physics, Faculty of Science, Kyoto University,
Kyoto 606-8502, Japan}
\author{Akira Ohnishi}
\affiliation{Yukawa Institute for Theoretical Physics, Kyoto University,
Kyoto 606-8502, Japan}
\author{Toru T. Takahashi}
\affiliation{Gunma National College of Technology, 
Gunma 371-8530, Japan}

\date{\today}
\pacs{11.15.Me, 12.38.Gc, 11.10.Wx, 25.75.Nq}

\begin{abstract}
We investigate possible  entropy production  in
Yang-Mills (YM) field theory by using a quantum distribution
 function called Husimi function $f_{\rm H}(A, E, t)$ for YM field, which
is given by a coarse graining of
 Wigner function and non-negative.
We calculate the Husimi-Wehrl (HW) entropy $S_{\rm HW}(t)=-{\rm Tr}f_H \log
f_H$
defined as an integral over the phase-space,
for which two adaptations of the test-particle method are used combined
with Monte-Carlo method.
We utilize the semiclassical approximation to obtain the time evolution of
the distribution functions of the YM field, which is known to show a
chaotic behavior
in the classical limit.
We also make a simplification of the multi-dimensional phase-space integrals
 by making a product ansatz for the Husimi function, which is found to give
 a 10-20 per cent over estimate of the HW entropy for a quantum system with a
few degrees of freedom.
We show that the quantum YM  theory does exhibit the entropy production, and
that the entropy production rate agrees with the sum of positive Lyapunov
exponents
or the Kolmogorov-Sinai entropy,
suggesting that the chaoticity of the classical YM field
causes the entropy production in the quantum YM theory.
\end{abstract}

\maketitle

{\it Introduction}.---
Thermalization or entropy production in an isolated quantum system is a long-standing problem. 
%
The entropy of a quantum system may be
given by von Neumann entropy $S_{\rm vN}=-\text{Tr}\,\rho\log\rho$ 
with $\rho$ being the density matrix~\cite{vonNeumann}, and 
taking into account that the time evolution of the quantum system is described by a unitary transformation $U(t)={\rm e}^{-iHt/\hbar}$, 
von Neumann entropy $S_{\rm vN}$
is shown to remain unchanged in time, which is an absurd consequence in contradiction to the reality.
One possible way to avoid this puzzle is to assume that there is no isolated quantum 
system because any quantum system is surrounded by the environment composed of  quantum fields described by, say, QED;
the partial trace with respect to the environment would lead to a
density matrix of a mixed state
due to the entanglement\cite{Zurek1991vd}.
For thermalization of a macroscopic quantum system, the old idea of von
Neumann
has been recently 
rediscovered,
and then a lot of related works and developments are being made
\cite{vonNeumann1929,Goldstein2010};
see \cite{Tasaki2015} and references cited therein.
It might be worth mentioning that the entanglement entropy of a quantum system may have a
geometrical interpretation as is clearly shown by Ryu-Takayanagi's formula \cite{Ryu:2006bv}.

In this work, we do not intend to
develop a master theory to describe thermalization or 
entropy production of a generic quantum system. 
Instead, we concentrate on
entropy production in quantum systems
whose classical counter parts are 
chaotic, and the semiclassical approximation is valid. There are many physical systems satisfying these characteristics \cite{CGC}: Among them, we have in mind the problem of the early thermaliztion 
in high-energy heavy-ion collisions  (see review \cite{Heinz02} and recent studies \cite{Baier01,Romatschke06,Muller06,Berges08,Iwazaki08,Fujii08,Fries08,
Fries09,Fujii09,Kunihiro10,Fukushima12,Epelbaum13,Iida13,Iida14,Tsutsui15-1,Tsutsui15-2,
Ruggieri15-1,Ruggieri15-2})
at the relativistic heavy-ion collider 
in the Brookhaven National Laboratory \cite{WPBRAHMS,WPPHENIX,WPPHOBOS,WPSTAR}
and the large hadron collider at CERN \cite{Muller12}.

Chaotic classical systems are characterized by the sensitive dependence
 of the trajectory on the initial condition, and
trajectories starting from adjacent initial conditions
with the difference $\delta X(0)$ in the phase
 space
deviate exponentially 
$|\delta X(t)|=\exp(\lambda t)|\delta X(0)|$ from each other:
The exponent $\lambda$ is called a  Lyapunov exponent.
Then one can readily imagine that 
the chaotic behavior makes the phase-space distribution $f(q, p)$ so complicated
to generate a finite amount of entropy via a coarse graining in the classical Hamiltonian system.
In this respect, it is interesting that the sum of positive Lyapunov exponents coincides with the 
Kolomogorov-Sinai entropy (see references in Ref.~\cite{KSentropy}) or the production rate of entropy \cite{Pesin}.   
Indeed, these have been demonstrated 
for a discrete classical system \cite{Latora1999},
where an explicit calculation of 
the Boltzmann-like 
entropy $S_\text{B}=-\text{Tr}f\log{f}$ 
was made
with the
distribution function $f(q,\,p)$ 
as obtained by a coarse graining of the phase space of the discrete system.

A natural  extension of the above interesting work to a quantum
system might be done with the application of the quantum mechanical distribution function, i.e., 
Wigner function $f_W(q,\,p)$ derived as a Weyl transform of the density matrix $\rho$ \cite{Wigner}.
However,  since $f_W$ is a mere Weyl transform of $\rho$, it can not describe 
an entropy production of a pure quantum system, even apart from the fact that
$f_W$ is not positive definite.


To circumvent this well known difficulty,
let us recall that one can not distinguish two phase space points
in a unit cell in quantum mechanics,
 and {\em smearing in the phase space volume of $(2\pi\hbar)^D$} 
may be allowed, where $D$ is the degrees of freedom (DOF) of the system.
As such a smeared distribution function, we adopt Husimi function $f_{\rm H}$ \cite{Husimi},
which is  obtained by a Gaussian smearing of  Wigner function 
and  semi-positive definite.
Then,
we can define the Boltzmann-like entropy 
in terms of $f_{\rm H}$ as
 $S_\text{HW}=-\text{Tr}\,f_{\rm H}\log{f_{\rm H}}$,
where $\text{Tr}$ means the integral over the phase space.
This entropy was first introduced and called the classical
entropy by  Wehrl \cite{Wehrl1978}, and
we call it Husimi-Wehrl (HW) entropy~\cite{KMOS,Tsukiji2015}.
In the previous work~\cite{Tsukiji2015},
the present authors examine thermalization of isolated quantum systems
by using the HW entropy evaluated in the semiclassical approximation.
It was shown that the semiclassical treatment
works well in describing the entropy-production process
of a couple of quantum mechanical systems
whose classical counter systems are known to be chaotic.
Two novel methods were also proposed to evaluate
the time evolution of the HW entropy,
the test-particle method and the two-step Monte-Carlo method, and it was demonstrated 
that the simultaneous application of the two methods ensures
the reliability of the results of the HW entropy at a given time.


In this article, we extend the previous work~\cite{Tsukiji2015} to
the Yang-Mills (YM) field,
which is known to be chaotic
and has macroscopic number of positive Lyapunov exponents~\cite{Kunihiro10}.
We investigate the possible entropy production
by constructing the Husimi function  and calculating
the HW entropy of the YM field in the semiclassical approximation.
The initial condition we adopt for the equation of motion (EOM) of the YM field  is motivated by
the early stage of relativistic heavy ion collisions \cite{CGC, Glasma}.

There is, however, a caveat against this simple prescription that works for quantum mechanical systems
with a few degrees of freedom because of the large number of the
degrees of freedom peculiar to the field theory.
Thus we also take a simple ansatz for the Husimi function, where
we construct it  by a product of Husimi function
for each degree of freedom,
although the classical EOM itself is solved numerically with the
fully included nonlinear
couplings. 
When applied to a quantum mechanical system with
two-degrees of freedom,
the ansatz gives 10 to 20 per-cent over estimate of 
the HW entropy. We also develop a novel efficient numerical method for calculating the 
HW entropy, which is a modification of the test-particle method.    
We show that the YM theory does exhibit the entropy production, and
find that the entropy production rate
agrees with the sum of positive Lyapunov exponents or KS entropy.
As long as we know, this is the first work to calculate the time dependence
of entropy in non-integrable field theory
except for the kinetic entropy shown in Ref.~\cite{Nishiyama}.

{\it Husimi-Wehrl entropy on the lattice}.---
We consider the $\text{SU}(N_c)$ YM field on a $L^3$ lattice.
In the temporal gauge, Hamiltonian in non-compact formalism is given by
\beq
H=\frac{1}{2}\sum_{x,a,i} E^a_i(x)^2 +\frac{1}{4}\sum_{x,a,i,j} F^a_{i j}(x)^2,
\eeq
with $F^a_{i j}=\partial_i A^a_j(x)-\partial_j A^a_i(x)+\sum_{b,c}f^{a b c}A^b_i(x)A^c_j(x)$.
$N_D=3L^3(N_c^2-1)$ is the total DOF.
We take the dimensionless gauge field $A$
and conjugate momentum $E$ normalized by the lattice spacing $a$
throughout this article.
The coupling constant $g$ is also included in the definition of $A$ and $E$.

The Husimi-Wehrl entropy of the YM field is obtained
as a natural extension of that in quantum mechanics
by regarding $(A(x),E(x))$ as canonical variables.
First, we define the Wigner function
(referred to as the Wigner functional~\cite{Mrowczynski1994})
in terms of $A(x)$ and $E(x)$,
\begin{align}
\fW[A,E;t]=&\int \frac{DA'}{(2 \pi \hbar)^{N_D}}
e^{i {E\cdot A'}/{\hbar}}
\nonumber\\
&\times\left\langle A+{A'}/{2}\mid\hat{\rho}(t)\mid{A-A'/2}\right\rangle.\label{fW}
\end{align}
where
$A\cdot E=\sum_{i,a,x} A^a_i(x) E^a_i(x)$ is the inner product.
The time evolution of the Wigner function is derived
from the von Neumann equation,
\beq
\frac{\partial}{\partial t}\fW[A,E;t]
=\frac{\partial H}{\partial A}\cdot\frac{\partial \fW}{\partial E}
-\frac{\partial H}{\partial E}\cdot\frac{\partial \fW}{\partial A}
+\mathcal{O}(\hbar^2)
.\label{EOM}
\eeq
In the semi-classical approximation,
we ignore $\mathcal{O}(\hbar^2)$ terms, then
$\fW$ is found to be constant along the trajectory
satisfying the classical equation of motion (EOM) (good review Ref.~\cite{Polkovnikov10}),
\beq
\dot{E}=-\frac{\partial H}{\partial A}, \, \dot{A}=\frac{\partial H}{\partial E}.\label{cEOM}
\eeq

Secondly, we introduce the Husimi function
as the smeared Wigner function with the minimal Gaussian packet,
\beq
\fH[A,E;t]=\int \frac{DA' DE'}{(\pi \hbar)^{N_D}} \mathrm{e}^{-\Delta (A-A')^2/\hbar-(E-E')^2/\Delta \hbar} \nonumber\\
\times\fW[A',E';t]\ ,\label{fH}
\eeq
where $\Delta$ is the parameter for the range of Gaussian smearing.
As in quantum mechanics,
Husimi function is semi-positive definite;$f_H[A,E;t]\ge0$,
and we define
the Husimi-Wehrl entropy 
as the Boltzmann's entropy or the Wehrl's classical entropy~\cite{Wehrl1978}
by adopting the Husimi function for the phase space distribution,
\beq
\SHW(t)=-\int \frac{DA DE}{(2\pi \hbar)^{N_D}}\fH[A,E;t]\log \fH[A,E;t].\label{SHW}
\eeq

{\it Numerical methods}.---
We calculate the time evolution of the HW entropy by two methods; test particle (TP) method and parallel test particle (pTP) method.
The TP method is developed in Ref.~\cite{Tsukiji2015}.
The pTP method, an alternative method for two-step Monte Carlo
method, requires less
numerical cost and gives almost the same results
as two-step Monte Carlo method.
We have demonstrated that the HW entropy in some quantum mechanical systems 
are successfully obtained in these two methods,
which are reviewed in the following.

In the test-particle method (TP),
we assume that the Wigner function is a sum of the delta functions,
\begin{align}
\fW[A,E;t]=&\frac{(2\pi\hbar)^\ND}{N_\mathrm{TP}}\sum_{i=1}^{N_\mathrm{TP}}
\delta^\ND(A-A_i(t))\,\delta^\ND(E-E_i(t)),\label{Eq:WignerTP}
\end{align}
where $N_{\rm TP}$ is the total number of the test particles.
The initial conditions of the test particles are
$(A_i(0),\,E_i(0))$\, $(i=1,\,2,\dots,\,\NTP)$,
which are chosen so as to well sample $\fW[A, E, 0]$.
The time evolution of the coordinates
$(A_i(t),E_i(t))$ 
is determined so that it reproduces the EOM for $\fW[A,
E, t]$, which is reduced to
the canonical equation of motion Eq.~(\ref{cEOM}) in the semiclassical approximation.

With the test-particle representation 
of the Wigner function, Eq.~\eqref{Eq:WignerTP},
the Husimi function is readily expressed as 
\begin{align}
\fH[A,E;t]=&\frac{2^\ND}{\NTP} \sum_{i=1}^{\NTP} 
e^{-\Delta (A-A_i(t))^2/\hbar-(E-E_i(t))^2/\Delta\hbar}
\ .\label{Eq:HusimiTP}
\end{align}
It is noteworthy that the 
Husimi function here is a smooth function
in contrast to the corresponding Wigner function in Eq.~\eqref{Eq:WignerTP}.

Substituting the Wigner function \eqref{Eq:HusimiTP} into Eq.~\eqref{SHW}, 
the HW entropy in the test-particle method is 
finally given as,
\begin{widetext}
\begin{align}
S_\mathrm{HW}^\mathrm{(TP,TTP)}
=&-\frac{1}{\NTP}
\sum_{i=1}^\NTP
\int \frac{D A D E}{(\pi\hbar)^\ND}\,
e^{-\Delta (A-A_i(t))^2/\hbar-(E-E_i(t))^2/\Delta\hbar}\log\left[\frac{2^\ND}{\NTP} \sum_{j=1}^{\NTP} 
e^{-\Delta (A-A_j(t))^2/\hbar-(E-E_j(t))^2/\Delta\hbar}\right]\nonumber\\
\simeq&-
\frac{1}{\NMC\NTP}\sum_{k=1}^\NMC\sum_{i=1}^\NTP
\log\left[
\frac{2^\ND}{\NTP} \sum_{j=1}^{\NTP} 
e^{-\Delta (\bar{A}_k+A_i(t)-A_j(t))^2/\hbar-(\bar{E}_k+E_i(t)-E_j(t))^2/\Delta\hbar}
\right].
\end{align}
\end{widetext}
Note here that the integral over $(A,E)_i$ for each $i$ has a support 
only around the positions of 
the test particles $(A_i(t),\,E_i(t))$
due to the Gaussian function,
and we can effectively perform the Monte-Carlo
integration in the second line.
With a set of random numbers $(\bar{A},\bar{E})_i$
with standard deviations of $\sqrt{\hbar/2\Delta}$ and $\sqrt{\hbar\Delta/2}$,
Monte-Carlo sampling point $(A,E)_i$ for each $i$ is generated as
$(A,E)_i=(A_i,E_i)+(\bar{A},\bar{E})_i$. 
Total sample number of
$(\bar{A},\bar{E})_i$ is denoted by $\NMC$ throughout this letter.

\begin{figure}[t]
\begin{center}
\includegraphics[width=85mm]{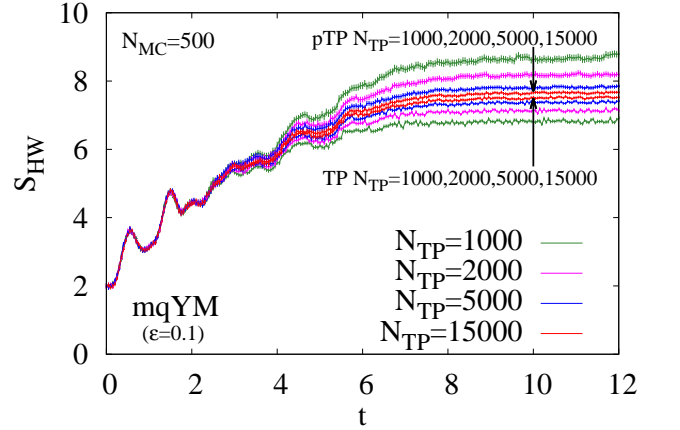}
\caption{Time evolution of Husimi-Wehrl entropy for the modified quantum Yang-Mills mechanics in TP and pTP methods.}
\label{mqYMe01}
\end{center}
\end{figure}

In the parallel test particle method, we make test particles in and out of logarithm 
independent, while they are same samples in TP.
Figure \ref{mqYMe01} shows the numerical results of the HW entropy
in the two dimensional quantum mechanical system, whose Hamiltonian is
given by
\beq
H=\frac{p^2_1}{2}+\frac{p^2_2}{2}+\frac{1}{2} q^2_1 q^2_2+\frac{\epsilon}{4}q^4_1+\frac{\epsilon}{4}q^4_2.
\eeq
This system is called a modified quantum Yang-Mills (mqYM) model in Ref.~\cite{Tsukiji2015}.
With increasing test-particle number,
the HW entropy is found to converge from below (above) in the TP (pTP) method,
then it is possible to give upper and lower limits of the entropy and to guess the converged value
by comparing the results in the two methods.

{\it Product ansatz and example in a 2-dim quantum mechanical system}.---
While the extension to the field theory on the lattice is straightforward,
the DOF are so large and numerical-cost demanding
in quantum field theories
that we need to invoke some approximation scheme in practical calculations.
We here adopt product ansatz to avoid this difficulty.

In the ansatz, we construct the Husimi function 
as a product of that for one degree of freedom,
\beq
\fH^\mathrm{(PA)}[A,E;t]=\prod^{N_D}_i \fH^{(i)} (A_i,E_i;t)\ ,
\label{Eq:PA}
\eeq
where $\fH^{(i)}=\int \prod_{j\not=i} dA_jdE_j/2\pi\hbar\, \fH[A,E;t]$.
By substituting this ansatz into Eq.~\eqref{SHW},
we obtain the HW entropy as a sum of the HW entropy for one degree of freedom;
\beq
\SHWPA
=-\sum^{N_D}_{i=1}\int\frac{dA_i dE_i}{2\pi \hbar}
\fH^{(i)}\,\log \fH^{(i)}\ .\label{SHWPA}
\eeq

The entropy estimated with the product ansatz gives the upper bound of the entropy, since it holds subadditivity.
The subadditivity of entropy is
expressed as
\begin{align}
S_1(\rho_1)+S_2(\rho_2) \geq S_{12}(\rho_{12})
\ ,
\end{align}
where $\rho_1=\int d \Gamma_2 \rho_{12}$ and $\rho_2=\int d \Gamma_1
\rho_{12}$ and $S_1$ and $S_2$ are subsystem entropies.
In this paper, we apply it to the Husimi function and the Husimi-Wehrl
entropy.
Thus obtained entropy $S_\mathrm{HW}^\mathrm{(PA)}$ gives the
upper bound of $S_\mathrm{HW}$ due to the subadditivity;
\begin{align}
S_\mathrm{HW}
\equiv S_{12...N_D}(\rho_{12...N_D})&\leq S_1(\rho_1)+S_{23...N_D}(\rho_{23...N_D})\nonumber\\
&\leq\sum^{N_D}_n S_n(\rho_n)=S_\mathrm{HW}^\mathrm{(PA)}
\ .
\end{align}

\begin{figure}[t]
\begin{center}
\includegraphics[width=85mm]{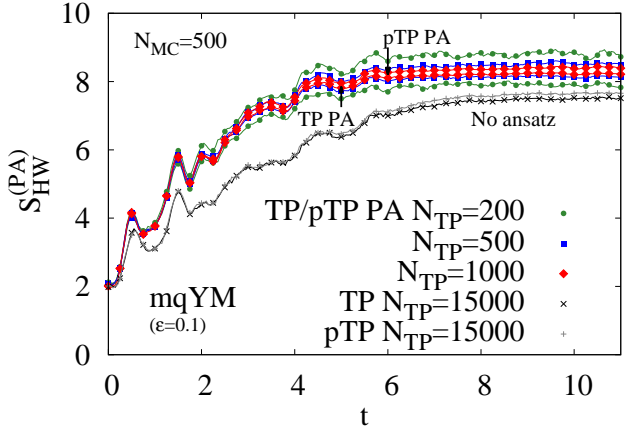}
\caption{The variation in the product ansatz in the two-dimensional quantum model with TP and pTP methods. 
The green circle, blue square and orange triangle lines are the time evolution of HW entropy with the product ansatz in TP and pTP methods where the numbers of test-particles are $\NTP=200, 500$ and $1000$, respectively. The gray cross and black x lines are the time evolution of the HW entropy in both methods without the product ansatz $\NTP=15000$.}
\label{PA}
\end{center}
\end{figure}

%
To check the variety we apply it to the mqYM model
previously discussed. 
Figure \ref{PA} shows the numerical results of the HW entropy
with the product ansatz ($\SHWPA$)
as well as the full entropy ($\SHW$), which can be found in Fig.~\ref{mqYMe01}.
While $\SHWPA$ slightly overestimates $\SHW$,
the difference is small enough to confirm entropy production.
%
The HW entropy with product ansatz is found to agree
with that without the ansatz within 10-20\% error
in a few dimensional quantum mechanical system.
We also find that numerical results with the ansatz
converge with smaller Monte Carlo samples, then it is much more efficient
from the view point of numerical-cost reduction.

{\it Entropy production in Yang-Mills field theory.}---
We apply the above-mentioned framework to the SU(2) Yang-Mills
field theory.
The initial condition of the Winger function is
set to be a Gaussian distribution,
$\fW[A,E;t=0]=2^{N_D}\exp(-\omega{A}^2/\hbar-E^2/\omega\hbar)$,
which corresponds to a coherent state.
The Wigner-function evolution is obtained by solving the classical EOM,
and the HW entropy with the product ansatz is calculated
by using the TP and pTP methods.
We take the parameter set, $\hbar=\omega=\Delta=1$.

\begin{figure}[t]
\begin{center}
\includegraphics[width=85mm]{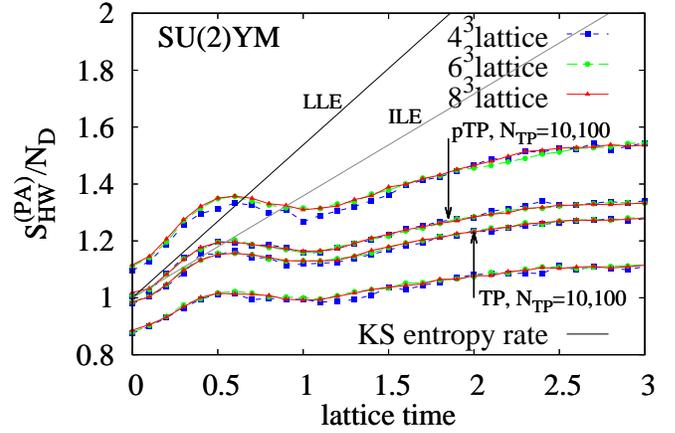}
\caption{The time evolution of Husimi-Wehrl entropy for the SU(2) Yang-Mills field theory on $4^3$ lattice in TP and pTP methods with the product ansatz. The blue square, green circle and red triangle lines are the HW entropy per one degrees of freedom on $4^3$lattice, $6^3$lattice and $8^3$lattice, respectively. The black and gray solid lines show growth rates of LLE and ILE, respectively, given in Ref.~\cite{Kunihiro10}.}
\label{su2YM}
\end{center}
\end{figure}

In Fig.~\ref{su2YM}, we 
show the time evolution
of the Husimi-Wehrl entropy per DOF
with the product ansatz in TP and pTP methods.
We find that the HW entropy per DOF is independent of the lattice size,
and the extensive nature of entropy is confirmed.
The dependence on the number of test-particle number
is the same as that in quantum mechanics;
With increasing Monte Carlo samples, $\SHWPA$ converges from below and
above in the TP and pTP methods, 
respectively.
The results in the TP and pTP   
methods approach each other with increasing $N_{\rm TP}$ and we can guess that the converged value lies between these curves.
After the oscillation around lattice time$=0.5$, the HW entropy increases in a monotonic way and its growth rate decreases.
This means that the collective motion in the earliest stage damps, and
the system approaches the equilibrium.
The entropy saturation occurs at $t\sim 3$ in the lattice unit in
the present setup.

The straight lines in Fig.~\ref{su2YM} show 
the Kolmogorov-Sinai entropy rate, which is given in Ref.~\cite{Kunihiro10}.
The upper and lower lines show the sum of positive local and intermediate
Lyapunov exponents (LLE and ILE), respectively.
LLE are obtained as the eigen values of the second derivative matrix
of the Hamiltonian,
and ILE show the exponential growth rate in some time duration.
Since the classical YM fields are conformal, the KS entropy rate
should be proportional to $\varepsilon^{1/4}$ where $\varepsilon$
is the energy per site.
The coefficients are evaluated in Ref.~\cite{Kunihiro10}
as $R_\mathrm{KS}^\mathrm{LLE}/L^3  \simeq 3\times \varepsilon^{1/4}$
and
$R_\mathrm{KS}^\mathrm{ILE}/L^3 \simeq 2\times \varepsilon^{1/4}$
for the sum of positive LLE and ILE (local and intermediate KS entropy rate),
respectively.
These findings show that the local KS entropy rate characterizes the growth rate of the HW entropy in the early time and the intermediate KS entropy rate agrees with the average entropy growth until the thermalization time.


{\it Conclusion}.---
In summary,
we have developed the numerical formulation to calculate the time evolution
of the Husimi-Wehrl (HW) entropy in Yang-Mills field theory
in the semi-classical approximation. 
This is a first work to calculate the time evolution of entropy in a
non-integrable field theory
except for the kinetic entropy shown in Ref.~\cite{Nishiyama}
as long as we know.
We have shown that the HW entropy is produced and the growth rates 
roughly agree with Lyapunov exponents.
It should be noted that the time reversal invariance is kept
in the present framework;
in the time evolution of the Wigner function
as well as in measuring the entropy.
The produced entropy mainly comes from the complexity
of the phase space distribution.

The entropy growth will have contribution to the thermalization process
in relativistic heavy ion collisions.
The set up in this article is motivated by
the initial stage dynamics in relativistic heavy ion collisions.
If we interpret the lattice spacing $a$ as the inverse saturation
scale $Q_s^{-1}=0.2$ fm/c at RHIC,
the lattice time scale is also normalized by the saturation scale $Q_s^{-1}$.
This means that the observed entropy-saturation time ($t\sim 3$ in the
lattice unit) is about $0.6$fm/c, which
might be the encouraging result for the early thremalization scenario which is suggested by the hydrodynamical simulation (review Ref.~\cite{Heinz02}).
For more realistic analysis,
we should choose the initial condition such as
the one given by the McLerran-Venugopalan model \cite{CGC}.




\section*{Acknowledgement}
%
This work was supported in part by 
the Grants-in-Aid for Scientific Research from JSPS
 (Nos.
 20540265, 
 23340067, 
 15K05079, 
 15H03663
),
the Grants-in-Aid for Scientific Research on Innovative Areas from MEXT
 (Nos. 23105713, 
       24105001, 24105008 
),
and
by the Yukawa International Program for Quark-Hadron Sciences.
T.K. is supported by the Core Stage Back Up program in Kyoto University.

\end{document}